\documentclass[12pt]{article}
\usepackage{amssymb,amsmath}
\include{epsf}

\newcommand{\eqn}[1]{(\ref{#1})}

\newbox\ncintdbox \newbox\ncinttbox
\setbox0=\hbox{$-$} \setbox2=\hbox{$\displaystyle\int$}
\setbox\ncintdbox=\hbox{\rlap{\hbox to
\wd2{\hskip-.125em\box2\relax \hfil}}\box0\kern.1em}
\setbox0=\hbox{$\vcenter{\hrule width 4pt}$}
\setbox2=\hbox{$\textstyle\int$}
\setbox\ncinttbox=\hbox{\rlap{\hbox to
\wd2{\hskip-.175em\box2\relax \hfil}}\box0\kern.1em}
\newcommand{\ncint}{\mathop{\mathchoice{\copy\ncintdbox}%
{\copy\ncinttbox}{\copy\ncinttbox}{\copy\ncinttbox}}\nolimits}

\newcommand{\mbf}[1]{{\boldsymbol {#1} }}
\newcommand{\mat}{{\bb M}} 
\newcommand{\id}{{1\!\!1}} 

\def\ii{{\,{\rm i}\,}}
\def\dd{{\rm d}}

\def\P{{\sf P}}

\def\a{{\sf a}}

\def\U{{\sf U}}
\def\V{{\sf V}}

\def\mPhi{{\mbf\Phi}}

\newcommand{\torus}{{\mathbb T}}
\newcommand{\circles}{{\mathbb S}}

\font\mybb=msbm10 at 12pt
\def\bb#1{\hbox{\mybb#1}}

\newcommand{\Tr}[1]{\:{\rm Tr}\,#1}
\def\e{{\,\rm e}\,}

\newcommand{\nn}{\nonumber \\}
\def\be{\begin{equation}}
\def\ee{\end{equation}}
\def\beq{\begin{equation}}
\def\eeq{\end{equation}}
\newcommand{\del}{\partial}
\def\bea{\begin{eqnarray}}
\def\eea{\end{eqnarray}}
\def\bd{\begin{displaymath}}
\def\ed{\end{displaymath}}

\newcommand{\atheta}{{{\cal A}_\theta}}

\newcommand{\athetan}{{{\cal A}_n}}

\makeatletter
\newdimen\normalarrayskip              
\newdimen\minarrayskip                 
\normalarrayskip\baselineskip
\minarrayskip\jot
\newif\ifold             \oldtrue            
\newcommand{\half}{{\textstyle{1\over 2}}}

\newcommand{\cc}{{\cal C}}

\newcommand{\ch}{{\cal H}}

\newcommand{\cs}{{\cal S}}

\newcommand{\IC}{{\mathbb C}}

\newcommand{\IM}{{\mathbb M}}
\newcommand{\IN}{{\mathbb N}}

\newcommand{\IR}{{\mathbb R}}
\newcommand{\IS}{{\mathbb S}}
\newcommand{\IT}{{\mathbb T}}
\newcommand{\IZ}{{\mathbb Z}}

\def\raw{\rightarrow}

\begin{document}
\begin{titlepage}
\begin{flushright}

\baselineskip=12pt
DSF--01--03\\ HWM--03--17\\ EMPG--03--16\\ hep--th/0309031\\
\hfill{ }\\
September 2003
\end{flushright}

\vspace{0.5 cm}

\begin{center}

\baselineskip=24pt

{\Large\bf A New Matrix Model for\\ Noncommutative Field Theory}

\baselineskip=14pt

\vspace{1cm}

{{\bf Giovanni Landi}$^{a,b,}$\footnote{\baselineskip=12pt Email: {\tt
landi@univ.trieste.it}}, {\bf Fedele
Lizzi}$^{b,c,d,}$\footnote{\baselineskip=12pt Email: {\tt
   fedele.lizzi@na.infn.it}} and {\bf
Richard J.~Szabo}$^{d,}$\footnote{\baselineskip=12pt Email: {\tt
R.J.Szabo@ma.hw.ac.uk}}}
\\[6mm]
$^a$ {\it Dipartimento di Scienze Matematiche, Universit\`a di
    Trieste\\ Via Valerio 12/b, I-34127 Trieste, Italia}
\\[6mm]
$^b$ {\it INFN, Sezione di Napoli, Napoli, Italia}
\\[6mm]
$^c$ {\it Dipartimento di Scienze Fisiche, Universit\`{a} di
Napoli {\sl Federico II}\\ Monte S.~Angelo, Via Cintia, 80126
Napoli, Italia}
\\[6mm]
$^d$ {\it Department of Mathematics, Heriot-Watt University\\
Scott Russell Building, Riccarton, Edinburgh EH14 4AS, U.K.}
\\[10mm]

\end{center}

\vskip 1 cm

\begin{abstract}

\baselineskip=12pt

We describe a new regularization of quantum field theory on the
noncommutative torus by means of one-dimensional matrix models. The
construction is based on the Elliott-Evans inductive limit
decomposition of the noncommutative torus algebra. The matrix
trajectories are obtained via the expansion of fields in a basis of
new noncommutative solitons described by projections and partial
isometries. The matrix quantum mechanics are compared with the usual
zero-dimensional matrix model regularizations and some applications
are sketched.

\end{abstract}

\end{titlepage}

\newpage

\setcounter{page}{2}

Field theories on noncommutative spaces possess many interesting
properties that are not shared by ordinary (commutative) quantum
field theories (see~\cite{NCrevs} for reviews). Most of these
novel features stem from the fact that the deformed algebra of
functions on spacetime is often best analysed when represented on
a separable Hilbert space, and noncommutative fields are most
naturally understood as operators. This implies that
noncommutative quantum field theories can be regulated and
constructively defined by means of finite-dimensional matrix
models. Furthermore, many of these field theories admit novel
solitonic solutions to their classical equations of motion which
have no counterparts in ordinary field theory. The solitons are
given as projections or partial isometries on the underlying
Hilbert space and correspond physically to configurations of
D-branes. In this Letter we will show that these two seemingly
disparate properties, the regularization and the presence of solitonic
configurations, are in fact intimately related.

A particularly tractable class of examples is provided by field
theories on the noncommutative torus, which may be constructed in
the following standard way~\cite{NCrevs}. Consider an ordinary
square two-torus $\IT^2$ with coordinate functions $U =\e^{2
\pi\ii x}$ and $V =\e^{2 \pi\ii y}$, where $x,y\in[0,1]$. By
Fourier expansion, the algebra $C^\infty(\IT^2)$ of complex-valued
smooth functions on the torus is made of power series of the form
\be\label{2ta}
a =\sum_{m,r=-\infty}^\infty a_{m,r}~U^m\,V^r
\end{equation}
with $\{a_{m,r}\}$ a complex-valued Schwartz function
on $\IZ^2$. Then, for a fixed real number $\theta$, the
algebra $\atheta =C^\infty(\IT^2_\theta)$ of smooth functions on the
noncommutative torus is the associative algebra made up of all elements
of the form \eqn{2ta}, but now the two unitary generators $U$ and $V$ no
longer commute but satisfy
\be\label{nct}
V\,U =\e^{2\pi\ii \theta}~U\,V~.
\end{equation}
There is a one-to-one correspondence between the noncommutative
torus $\atheta$ and functions on the commutative torus
$C^\infty(\IT^2)$ given by the Weyl map
$\Omega:C^\infty(\IT^2)\to\atheta$ and its inverse, the Wigner map
$\Omega^{-1}$ which is defined on algebra elements of the form
(\ref{2ta}) by
\be
\Omega^{-1}(a)(x,y)=
\sum_{m,r=-\infty}^\infty a_{m,r}~\e^{-\pi\ii m\,r\,\theta}~
\e^{2\pi\ii(m\,x +r\,y)} \ .
\label{wigmap}\end{equation}
This correspondence, which is an isomorphism of vector spaces but
clearly does not preserve the multiplication, can be used to deform
the commutative product on the algebra $C^\infty(\IT^2)$ into a
noncommutative star-product in the usual way.

The construction of field theories on $\atheta$ now proceeds by
introducing the normalized, positive definite trace denoted by
$\ncint :\atheta \raw \IC$ given on elements of the form
(\ref{2ta}) by
\be
\ncint a:= a_{0,0}~.
\label{ncintadefa00}\end{equation}
This definition gives the integration of the corresponding
Wigner function (\ref{wigmap}) over the ordinary torus
$\IT^2$. Derivatives on $\atheta$ are determined by means of two
commuting linear derivations $\del_\mu$, $\mu=1,2$ defined by
\beq
\del_1U= 2\pi\ii U~, ~~~\del_1V=0~~~;~~~
\del_2U=0 ~, ~~~\del_2V= 2\pi\ii V ~.
\label{t2act}\end{equation}
The invariance of the trace is the statement that $\ncint \del_\mu a
=0$ for any $a\in \atheta$.
The generic situation is that there is a set of fields,
which we denote collectively by $\Phi$, with Lagrangian density $\cal
L$, all of which are elements of $\atheta$. Then the action for a
noncommutative field theory on the two-torus is given schematically by
\be
S=\ncint{\cal L}[\Phi,\partial_\mu\Phi] \ .
\label{NCstrfield}
\end{equation}

Such field theories are well-known to be regulated by means of
finite-dimensional matrix models~\cite{AMNS,LLS} via the observation
that any irrational number $\theta$ can be treated as a limit
\be
\theta=\lim_{n\to\infty}\,\theta_n \ , ~~ \theta_n:=\frac{p_n}{q_n}
\label{thetalim}\end{equation}
of rational numbers $\theta_n$ in a definite way by using continued
fraction expansions. At each finite level $n$, the integers $p_n$ and
$q_n>0$ are relatively prime, and with them one introduces the
$q_n\times q_n$ clock and shift matrices
\be
\cc_{q_n}=\left({\begin{array}{lllll}
1& & & & \\ &\e^{\frac{2\pi\ii p_n}{q_n}}& & & \\
& &\e^{\frac{4\pi\ii p_n}{q_n}}& & \\& &
&\ddots& \\ & & & &
\e^{\frac{2(q_n-1)\,\pi\ii p_n}{q_n}}
\end{array}}\right) \ , ~~
\cs_{q_n}=\left({\begin{array}{lllll}
0&1& & &0\\ &0&1& & \\
& &\ddots&\ddots& \\
& & &\ddots&1\\ 1& & & &0\end{array}}\right)
\label{torfuz}\end{equation}
which are unitary and traceless (since $\sum_{k=0}^{q_n-1}\e^{2k\,\pi\ii
  p_n/q_n}=0$), satisfy
  $(\cc_{q_n})^{q_n}=(\cs_{q_n})^{q_n}=\id_{q_n}$, and obey the
  commutation relation
\be
\cs_{q_n}\,\cc_{q_n} = \e^{2\pi\ii p_n/q_n}~\cc_{q_n}\,\cs_{q_n} \ .
\label{clockshiftcommrel}\end{equation}
Since $p_n$ and $q_n$ are relatively prime, the matrices \eqn{torfuz}
generate the finite dimensional algebra $\IM_{q_n}(\IC)$ of $q_n\times
q_n$ complex matrices. The matrix algebra generated by $\cc_{q_n}$ and
$\cs_{q_n}$ is also commonly referred to as the fuzzy torus.

The noncommutative field theory with action functional
(\ref{NCstrfield}) may then be approximated by the zero-dimensional
matrix model with action
\be
S_n^{(0)}=\Tr\,{\cal
  L}\big[\pi_n(\Phi)\,,\,d_\mu\,\pi_n(\Phi)\,d_\mu^\dag-
\pi_n(\Phi)\big]
\label{Sn0Dapprox}\end{equation}
defined in terms of the surjective map $\pi_n:{\cal
  A}_\theta\to\IM_{q_n}(\IC)$ which is given on elements of the
  form (\ref{2ta}) by
\be\label{calsur1}
\pi_n(a)=\sum_{m,r=-\infty}^\infty a_{m,r}~\left(\cc_{q_n}\right)^m
\,\left(\cs_{q_n}\right)^r~.
\end{equation}
In (\ref{Sn0Dapprox}), $\Tr$ denotes the usual $q_n\times q_n$ matrix
trace, while the canonical derivatives in (\ref{t2act}) are
approximated by the inner automorphisms generated by the matrices (for
odd $q_n$)
\be
d_1=\left(\cc_{q_n}^\dag\right)^{(q_n+1)/2} \ , ~~
d_2=\left(\cs_{q_n}\right)^{(q_n+1)/2}
\label{dmu0D}\end{equation}
which are defined directly in terms of the clock and shift matrices
(\ref{torfuz}). These matrix models are just field theories on the
fuzzy torus and are directly related to the
lattice regularization of noncommutative field theory (with $q_n$ the
size of the lattice)~\cite{AMNS}. The precise definition of the large
$n$ limit, required to reproduce the original continuum dynamics, is
described in~\cite{LLS}.

On the other hand, noncommutative soliton configurations
corresponding to D-branes wrapped around the torus $\IT^2$ have
been constructed in~\cite{T2solitons}. Projection solitons based
on the Powers-Rieffel~\cite{ri81} and Boca~\cite{bo} projections
on the noncommutative torus yield soliton configurations with the
correct K-theory charges. Here we describe new soliton
configurations which are similar in form to those built on the
Powers-Rieffel projection, but which possess some qualitatively
different properties. From these projections we also construct new
complex soliton fields corresponding to partial isometries in an
appropriate representation of the noncommutative torus algebra.
These solitons all come in sequences, based on the limiting
process~(\ref{thetalim}), and generate subalgebras of $\atheta$
which are naturally isomorphic to two copies of the algebra of
matrix-valued functions on a circle. We thereby obtain a
regularization of noncommutative fields which on the one hand is
given by expansions of the fields in soliton configurations, and
which on the other hand allows us to approximate the quantum field
theory with action~(\ref{NCstrfield}) by means of a {\it
one-dimensional} matrix model. The construction is based on the
Elliott-Evans inductive limit decomposition of the noncommutative
torus~\cite{ee}. From a mathematical perspective this
approximation is superior to those provided by zero-dimensional
matrix models, because there the noncommutative torus algebra
$\atheta$ can never be realized as an inductive limit of
finite-dimensional matrix algebras~\cite{LLS}, while in the
present case $\atheta$ is directly the inductive limit of matrix
algebras over circles. From a physical perspective, this means
that the ensuing matrix quantum mechanics has more tractable
features than the lattice matrix models. The technical details of
these constructions, as well as applications of the new matrix
model regularization, will be presented elsewhere~\cite{inprep}.

Our matrix models are based on the construction of new solitonic
configurations on the noncommutative torus, which we will describe
first. Following~\cite{ee}, we define two generalized families of projections
$\{\P_n\}_{n\geq1}$ and $\{\P_n'\}_{n\geq1}$ which are related to the
even and odd order approximants of the noncommutativity parameter in
(\ref{thetalim}). For each $n\in\IN$ we define two Powers-Rieffel type
projections by
\bea
\P_n & = & V^{-q_{2n-1}}\,\Omega(g_n)+\Omega(f_n)+\Omega(g_n)\,V^{q_{2n-1}} \ ,
\nn
\P_n' & = &U^{q_{2n}}\,\Omega(g'_n)+\Omega(f'_n)+\Omega(g'_n)\,U^{-q_{2n}} \ ,
\label{projdef}\end{eqnarray}
where $f_n$ and $g_n$ (resp. $f_n'$ and $g_n'$) are elements of the
subalgebra $C^\infty(\IS^1)\subset C^\infty(\IT^2)$ generated by
$\Omega^{-1}(U)$ (resp. $\Omega^{-1}(V)$). To simplify notation in the
following, we will suppress the
subscript $n$ on the functions $f$, $g$, $f'$ and $g'$, and the
subscripts $2n$ and $2n-1$ on the integers $p$ and $q$. To distinguish
$q_{2n}$ from $q_{2n-1}$ we will denote the former integer by $q$ and
the latter one by $q'$, and similarly for $p$. We shall further simply
write $\P=\P_n$ and $\P'=\P'_n$. The large $n$ limit is then
understood as the limits $q\to\infty$ and $q'\to\infty$.

As for the Powers-Rieffel projection, the real functions $f$ and
$g $ in~(\ref{projdef}) are ``bump'' functions which are chosen so
that the projection $\P$ has certain special properties. Define
\be
\beta=p'-q'\,\theta \ , ~~ \beta' = q\,\theta -p ~.
\end{equation}
{}From the continued fraction decomposition of (\ref{thetalim}) we then
have the relation~\cite{LLS,inprep}
\be\label{diph}
q\,\beta + q'\,\beta' = 1 ~.
\end{equation}
With an appropriate choice for the bump functions, the rank of $\P$ is
$\ncint \P = \beta$, while its monopole charge (Chern number) is
$Q(\P)=-q'$. Thus the projection $\P$ in (\ref{projdef}) represents a
soliton configuration carrying $({\rm D2},{\rm D0})$ brane charges
$(p',-q')$. In a completely analogous manner one finds $\ncint \P' =
\beta'$ and $Q(\P') = q$, and thus the projection $\P'$ has brane
charges $(-p,q)$.

\begin{figure}[htb]
\epsfxsize=2.5 in
\bigskip
\centerline{\epsffile{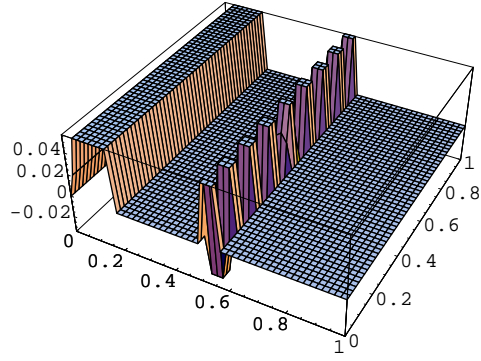}}
\caption{\baselineskip=12pt {\it The soliton field configuration
     corresponding to the projection operator $\P$ on the
     noncommutative torus. The noncommutativity parameter is taken to
     be the inverse of the golden mean, $\theta=\frac{2}{\sqrt5+1}$,
     while the approximants are chosen as $\theta_{2n}=\frac35$ and
     $\theta_{2n-1}=\frac58$. The vertical axis is the Wigner function
     $\Omega^{-1}(\P)(x,y)$ and the horizontal plane is the
     $(x,y)$-plane.}}
\bigskip
\label{zawinul3}\end{figure}

The Wigner function (\ref{wigmap}) on $\torus^2$ corresponding to the
projection $\P$ is easily computed, 
\be
\Omega^{-1}(\P)(x,y)=f(x)+
2\cos\left(2\pi\,q'\,y\right)\,g\left(x-\half \, q'\,\theta\right)\ .
\label{OmegaP}
\end{equation}
With the required bump functions, the soliton field (\ref{OmegaP}) has
a shape which is depicted in fig.~\ref{zawinul3}. Note that each
physical field configuration (\ref{OmegaP}) is concentrated in two
regions, each of which is localized along the $x$-cycle of the torus
but extended along the $y$-direction. It therefore defines lumps that have
strip-like configurations, unlike the standard point-like
configurations of GMS solitons on the noncommutative
plane~\cite{GMS}. The first lump has a smooth locus of points and
strip area depending on both the D-brane charge $\beta$ and the
monopole charge $q'$. The second lump contains a periodically spiked
locus of support points, with period $q'$. The spiking exemplifies the
UV/IR mixing property~\cite{MvRS} that generic noncommutative fields
possess, in that the size of the configuration decreases as its
oscillation period (the monopole charge) grows.

Given the projection $\P$, we ``translate'' it by the (outer)
automorphism $\rho$ defined by
\be
\rho(U)=\e^{2\pi\ii p/q}~U \ , ~~ \rho(V)=V \ .
\label{rhoautodef}
\end{equation}
The corresponding Wigner function (\ref{OmegaP}) is translated
accordingly along the $x$-cycle of the torus,
$\Omega^{-1}(\rho(\P))(x,y)=\Omega^{-1}(\P)(x+p/q,y)$. By
repeatedly applying $\rho$ we can define new projections
\bea
\P^{ii}~:=~\rho^{i-1}(\P)
&=&V^{-q'}\,\Omega\Big(g\big(x+(i-1)\,p/q\big)\Big)\nn &&
+\,\Omega\Big(f\big(x+(i-1)\,p/q\big)\Big)+
\Omega\Big(g\big(x+(i-1)\,p/q\big)\Big)\,V^{q'}
\label{transe}
\end{eqnarray}
for $i=1,\dots,q$. From $\rho^q ={\rm id}$ it follows that
$\P=\P^{11}=\P^{q+1,q+1}$. Moreover, one checks that
the elements \eqn{transe} form a system of mutually
orthogonal projection operators,
i.e. $\P^{ii}\,\P^{jj}=\delta_{ij}\,\P^{jj}$. As the notation
suggests, these projections are the diagonal elements of a basis for a
certain matrix subalgebra of $\atheta$ which we are now going to
describe.

\begin{figure}
\bigskip
\centerline{\epsfxsize=2.5 in\epsffile{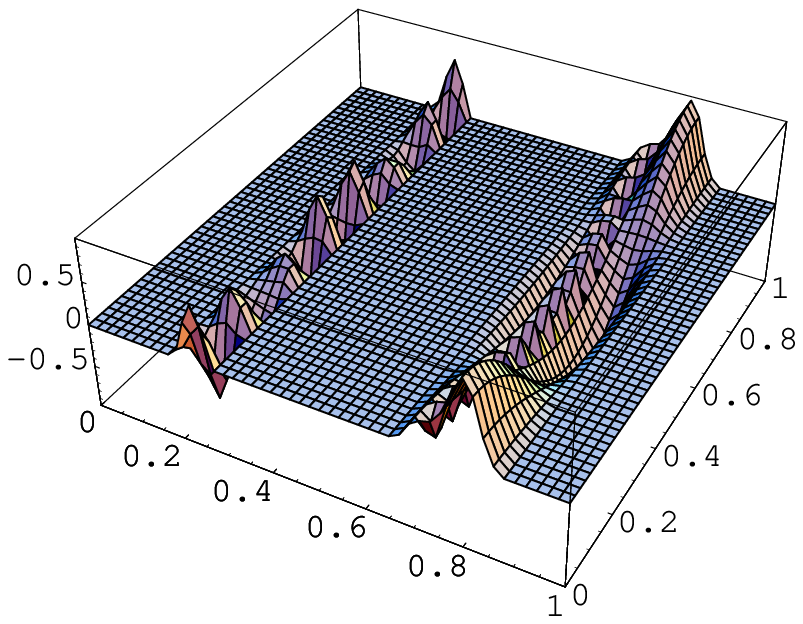}\hskip 1in
\epsfxsize=2.5 in
\epsffile{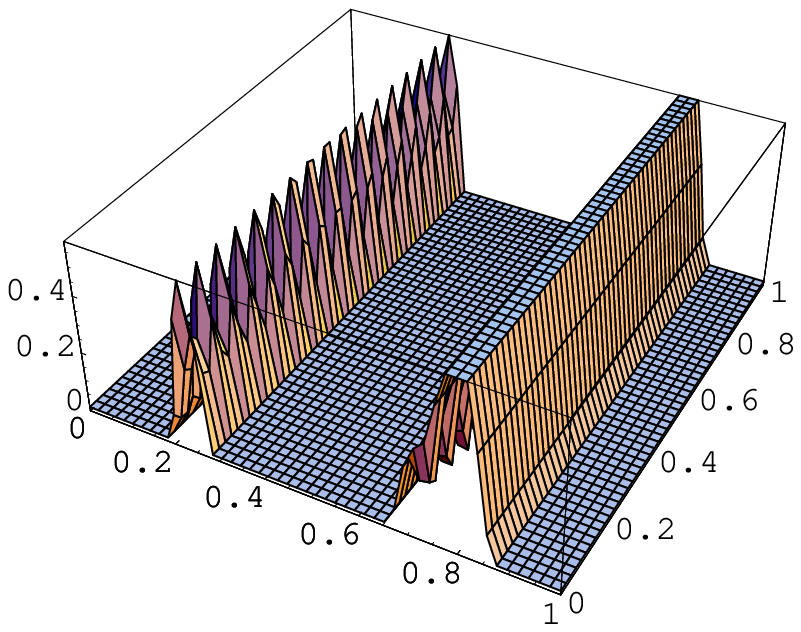}}
\caption{\baselineskip=12pt {\it The soliton field configuration
     corresponding to the operator $\Pi^{21}$ on the
     noncommutative torus. Displayed are its real part (left) and
     modulus (right). Its imaginary part has shape which is qualitatively the
     same as that of its real part. Parameter values and axes are as in
     fig.~\ref{zawinul3}.}}
\bigskip
\label{zawinul4}\end{figure}

We will implicitly represent the noncommutative torus algebra on the GNS
representation space $\ch = L^2(\atheta\,,\,\ncint\,)$, which is just
the completion of the algebra $\atheta$ itself in the norm defined by
the trace (\ref{ncintadefa00}). Let $\ch_i\subset\ch$ be the range of the
projection $\P^{ii}$, on which it acts as the identity $\id$, while
for $j\neq i$ one has $\ch_i\subset\ker(\P^{jj})$.
We will also need another set of operators which map one
subspace into another, as they will be the off-diagonal elements of
the matrix algebra basis. For this, we consider the operator
\be
\Pi^{21}:=\P^{22}\,V\,\P^{11} \ .
\label{Pi21}
\end{equation}
This operator is a mapping from $\ch_1$ to $ \ch_2$, but it is not
an isometry, i.e. $(\Pi^{21})^\dag\,\Pi^{21}\neq\id$. Therefore we
introduce a related {\it partial} isometry $\P^{21}$, i.e.\ an
operator for which $(\P^{21})^\dag\,\P^{21}$ and
$\P^{21}\,(\P^{21})^\dag$ are projection operators. Such an
operator is naturally given by the partial isometry appearing in
the polar decomposition
\be
\Pi^{21}:=\P^{21}\,\left|\Pi^{21}\right| ~, ~~~~~\, \,
\left|\Pi^{21}\right|=\sqrt{\,\left(\Pi^{21}\right)^\dag\, \Pi^{21} }~ ,
\label{Pi21polar}
\end{equation}
which is well-defined since the operator (\ref{Pi21}) is bounded.
The decomposition~(\ref{Pi21polar}) is understood as an equation
in the representation of the algebra $\atheta$ on the Hilbert
space $\cal H$, so that $\P^{21}\in\atheta$. The physical
significance of such an operator is that it is unitary in the
orthogonal complement to a kernel and cokernel, and hence will
produce localized solitons (in the Wigner representation). The
operator $\Pi^{21}$ and the partial isometry $\P^{21}$ converge to
each other in norm in the large $n$ limit~\cite{ee}. A
straightforward calculation gives the Wigner function on
$\torus^2$ corresponding to the operator~(\ref{Pi21}) in terms of
the periodic bump functions. Its shape is plotted in
fig.~\ref{zawinul4}. Again, the multi-soliton aspect is apparent,
with smooth and periodically spiked support loci. Note that while
the modulus of the function expectedly displays the characteristic
strips of projection solitons, the lumps of its real and imaginary
parts are point-like configurations.

Using (\ref{Pi21}) and (\ref{Pi21polar}), we define translated
partial isometries analogously to~\eqn{transe} as
\be
\P^{i+2,i+1}:=\rho^i\left(\P^{21}\right) \ , ~~ i=1,\ldots, q-2 \ .
\label{ejkdef}
\end{equation}
Finally, we also define $\P^{ji}:=\left(\P^{ij}\right)^\dag$.
The important fact is that for the
operators (\ref{transe}) and (\ref{ejkdef}) which we have defined,
there is a set of relations~\cite{ee,inprep}
\be
\P^{ij}\,\P^{kl}=\delta_{jk}\,\P^{il} \ . \label{matrixmult}
\end{equation}
These relations suggest the definition of $q^2$ operators $\P^{ij}$,
$1\leq i,j\leq q$. The remaining cases ($j\neq i$ and $j\neq i\pm1$) are
{\it defined} by (\ref{matrixmult}). For example,
$\P^{13}:=\P^{12}\,\P^{23}$, and so on. The newly defined operators
$\P^{ij}$ obtained in this way are partial isometries which are maps
from $\ch_j$ to $\ch_i$, i.e. elements of
$\P^{ii}\,\atheta\,\P^{jj}$. For the collection of all of them
$\{\P^{ij}\}_{1\leq i,j\leq q}$, the relation~\eqn{matrixmult}
    holds. In this way we can complete the sets of operators
    (\ref{transe}) and (\ref{ejkdef}) into a system of
matrix units which generate a $q\times q$ matrix algebra.

There is, however, a caveat. The operators $\P^{i+2,i+1}$
in~\eqn{ejkdef} are only defined for $i\leq q-2$, and this is in fact
sufficient to define all of the $\P^{ij}$ using (\ref{matrixmult}), including
\be
\P^{1q}:=\P^{12}\,\P^{23}\cdots\P^{q-1,q} \ . \label{defe1q}
\end{equation}
On the other hand, we can also define
\be
\tilde\P^{1q}:=\rho^{q-1}\left(\P^{21}\right) \ . \label{defetilde1q}
\end{equation}
For the $q\times q$ matrix algebra to close, it would be necessary
that the two operators defined by~\eqn{defe1q} and~\eqn{defetilde1q}
coincide. This is \emph{not} the case. However, although they are not
identical, both of these operators are isometries from $\ch_q$ to
$\ch_1$. As a consequence, they are related by an operator $z$ which
is unitary on $\ch_1$, i.e. a unitary element of
$\P^{11}\,\atheta\,\P^{11}$, and which is therefore a partial isometry
on the full Hilbert space $\ch$. We therefore have
\be
\tilde\P^{1q}:=z~\P^{1q} \ .
\label{tildePz}\end{equation}
All this means that the matrix units $\P^{ij}$, {\it along with} the
partial isometry $z$, close a subalgebra of $\atheta$, in which,
using (\ref{matrixmult}) and (\ref{tildePz}), a generic element is a
complex linear combination of the form
\be \label{Az}
\a(z)=\sum_{k=-\infty}^\infty~\sum_{i,j=1}^qa_{ij;k}~z^k
\,\P^{ij}
\end{equation}
with $a_{ij;k}$ of sufficiently rapid descent as $k\to\infty$. By
regarding $z:=\e^{2\pi\ii\tau}$, $\tau\in[0,1)$ as the unitary
generator of a circle $\IS^1$, this subalgebra is (naturally
isomorphic to) the algebra $\mat_q(C^\infty(\IS^1))$ of $q\times q$
matrix-valued functions on the circle. The identity element of this
subalgebra is $\id_q=\sum_{i=1}^q\P^{ii}$. From the above definitions
it follows that the trace of the matrix units is given by
\be
\ncint\P^{ij}=\beta\,\delta_{ij} \ ,
\label{Pijtrace}
\end{equation}
and, in particular, the identity element has trace
$\ncint\id_q=q\,\beta$.

Similarly (modulo a gauge transformation, i.e. conjugation by a unitary
element in the algebra $\atheta$), starting from the projection $\P'$ we can
build an orthogonal set of matrix units
$\P^{\prime\,i'j'}$, $1\leq i',j'\leq q'$ which again close a
$q'\times q'$ matrix algebra up to a partial isometry
$z':=\e^{2\pi\ii\tau'}$, $\tau'\in[0,1)$. For each integer $n$,
one thereby generates an algebra which is isomorphic to a matrix
algebra
\be
\athetan~\cong~\mat_{q}\left(C^\infty(\circles^1)\right)\oplus
\mat_{q'}\left(C^\infty(\circles^1)\right) \ , \label{An}
\end{equation}
with the direct sum arising from the orthogonality of the two sets of
matrix generators based on the projections $\P$ and $\P'$ (to 
obtain this orthogonality one needs the aforementioned gauge
transformation). The important point here is that
$\athetan$, being constructed out of elements of the noncommutative torus, is
a {\it subalgebra} of
$\atheta$. The fact that this subalgebra approximates the
noncommutative torus resides in the property that for each element
$a\in\atheta$, it is possible to construct a corresponding element
$\a\in\athetan$ which approximates it in norm.
For this, at each level $n$ we introduce the roots of unity
$\omega=\e^{2\pi\ii\theta_{2n}}$ and
$\omega'=\e^{2\pi\ii\theta_{2n-1}}$ with $\omega^q=1=(\omega')^{q'}$,
and define
\bea
\U&=&\left(\,\sum_{i=1}^{q}\omega^{i-1}~\P^{ii}\right)\oplus
\left(\,\sum_{i'=1}^{q'-1}\P^{\prime\,i',i'+1}+z'\,
\P^{\prime\,q',1}\right)
{}~=~\begin{pmatrix}\,{\cal C}_{q}& 0 \cr
0 &{\cal S}_{q'}(z'\,)\,\cr\end{pmatrix} \ , \nn
&~& \nn
\V&=&\left(\,\sum_{i=1}^{q-1}
\P^{i,i+1}+z\,\P^{q,1}\right)\oplus\left(\,
\sum_{i'=1}^{q'}(\omega')^{i'-1}~\P^{\prime\,i'i'}\right)
{}~=~\begin{pmatrix}\,{\cal S}_{q}(z)& 0 \cr 0 &{\cal C}_{q'}\,\cr
\end{pmatrix} \ ,
\label{defUVn}\end{eqnarray}
where ${\cal C}_q$ is the $q\times q$ unitary clock matrix in
\eqn{torfuz}, while for any $z\in\circles^1$, ${\cal S}_q(z)$ is the
generalized $q\times q$ unitary shift matrix
\be
{\cal S}_q(z)=\left(
\begin{matrix}\,0&1& & &0\,\cr &0&1& & \cr
   &  &\ddots&\ddots& \cr & & &\ddots&1\,\cr\,z& & & &0\,\cr
\end{matrix}\right)
\label{zshift}
\end{equation}
with ${\cal S}_q(z)^q=z\,\id_q$. As already mentioned, the
matrices ${\cal C}_q$ and ${\cal S}_q={\cal S}_q(1)$ form a basis for the
finite-dimensional algebra  $\IM_q(\IC)$ of $q\times q$
complex-valued matrices. Then, the matrices (\ref{defUVn}) generate the
infinite-dimensional algebra (\ref{An}) of matrix-valued functions on
two circles. Moreover, $\U$ and $\V$ have a commutation relation which
approximates the one (\ref{nct}) of $U$ and $V$,
\be
\V\,\U={\mbf\omega}~\U\,\V
\label{UnVnomegan}
\end{equation}
with
\be
{\mbf\omega}=\omega\,\sum_{i=1}^{q}\P^{ii}
{}~\oplus~\omega'\,\sum_{i'=1}^{q'}\P^{\prime\,i'i'}
=\begin{pmatrix}\,\omega\,\id_{q}& 0 \cr 0
    &\omega'\,\id_{q'}\,\cr\end{pmatrix} \ .
\end{equation}
In all of these expressions we have stressed the important double
interpretations of the generators. The first equality emphasizes
that they are still elements of the algebra $\atheta$ (i.e. they
are linear combinations of solitons on the noncommutative torus),
while the second reminds that they are elements of a matrix
algebra (i.e. they are matrix-valued fields on two circles).

In the limit $n\to\infty$, both the even and odd approximants $\theta_{2n}$ and
$\theta_{2n-1}$ converge to $\theta$, while $q,q'\to\infty$, and the generators
$\U$ and $\V$ of $\athetan$ converge in norm to the generators $U$
and $V$ of $\atheta$~\cite{ee}. For each $n$ one constructs a
projection $\gamma_n:\atheta\to\athetan$, similar
to~(\ref{calsur1}), defined on elements of the form (\ref{2ta}) by
\be
\a\oplus\a':=\gamma_n(a)  =\sum_{m,r=-\infty}^\infty a_{m,r}~\U^m\,
\V^r \ .\label{Gammanhomo}
\end{equation}
It is crucial that for any element $a\in\atheta$, its projection
$\gamma_n(a)$ is very close to it in norm~\cite{inprep}, i.e.\ for
$n$ large enough, to each element of $\atheta$ there always
corresponds an element of the subalgebra $\athetan$ to within an
arbitrarily small radius. Only the information about the higher
momentum modes $a_{m,r}$ of the expansion of $a$ is lost (i.e. for
$m,r>q\,q'$), and these coefficients are small for Schwartz
sequences. Hence the approximation for large $n$ is good. In what
follows we will encode the noncommutativity of the algebra
$\athetan$ by using the usual matrix multiplication of functions
on a circle $\circles^1$. We then obtain an expansion of
noncommutative fields in terms of D-brane solitons on the torus
$\torus^2$. The remarkable fact about this soliton expansion is
that it leads to a description of the dynamics of a noncommutative
field in a very precise way in terms of a one-dimensional matrix
model, whose (inductive) limit reproduces {\it
  exactly} the original continuum dynamics~\cite{inprep}.

We now use the mapping (\ref{Gammanhomo}) onto the approximating
subalgebra $\athetan$ in (\ref{An}) to build a matrix quantum
mechanics which regulates a generic field theory
(\ref{NCstrfield}) on $\atheta$. For this, we need to transcribe
the integration and derivatives on $\atheta$ into operations
intrinsic to the matrix model, as was done in (\ref{Sn0Dapprox}).
Let us start with the canonical trace. On elements of $\atheta$
the trace (\ref{ncintadefa00}) is determined through the
definition $\ncint U^m\,V^r:=\delta_{m0}\,\delta_{r0}$. To
determine the trace of corresponding elements of $\athetan$, we
note that because of (\ref{Pijtrace}), traces of powers
$\U^m\,\V^r$ of the generators (\ref{defUVn}) vanish unless the
corresponding powers of both the clock and shift operators are
proportional to the identity elements $\id_{q}$ or $\id_{q'}$,
which happens whenever $m$ and $r$ are arbitrary integer multiples
of $q$ or $q'$. From (\ref{defe1q})--(\ref{tildePz}) it then
follows that
\be
\ncint\U^m\,\V^r=\sum_{k=-\infty}^\infty\left(q\,
\beta\,\delta_{m\,,\,q\,k}\,\delta_{r0}+q'\,
\beta'\,\delta_{m0}\,\delta_{r\,,\,q'\,k}\right) \ .
\label{ncintUnVnpowers}
\end{equation}
In the large $n$ limit, by using (\ref{diph}) we see that the
trace $\ncint\gamma_n(a)$ is therefore well approximated by
$a_{0,0}$, since the correction terms $a_{q\,k\,,\,0}$ and
$a_{0\,,\,q'\,k}$ for $k\neq0$ are then small for Schwartz
sequences. It is now clear how to rewrite $\ncint\gamma_n(a)$ in
terms of operations which are intrinsic to the matrix algebras
(\ref{An}). The trace of the unitaries $z=\e^{2\pi\ii\tau}$ and
$z'=\e^{2\pi\ii\tau'}$ can be reproduced on functions on
$\circles^1$ by integration over the circle, while the trace of
the matrix degrees of freedom are ordinary $q\times q$ and
$q'\times q'$ matrix traces $\Tr$ and $\Tr'$, respectively,
accompanied by the appropriate normalizations $q\,\beta$ and
$q'\,\beta'$. It then follows that the trace of a generic element
in (\ref{Gammanhomo}) can
be written solely in terms of matrix quantities as
\be
\ncint\a\oplus\a'=\beta\,\int\limits_0^1
\dd\tau~\Tr\,\a(\tau)+\beta'
\,\int\limits_0^1\dd\tau'~\Tr'\,\a'(\tau'\,) \ .
\label{ncintazzprime}
\end{equation}

The definition of the equivalent $\nabla_\mu$, $\mu=1,2$ of the
derivations~\eqn{t2act} on $\athetan$ is
somewhat more involved. For this, let us look more closely at the map
$a\mapsto\gamma_n(a)$ defined in~(\ref{Gammanhomo}), and
express it as a power series expansion
\bea
\a\oplus\a'(z,z'\,)&=&\sum_{i,j=1}^q~\sum_{k=-\infty}^\infty\,
\alpha_{i+q/2,j;k}~z^k\,\left({\cal C}_{q}\right)^i\,
\big({\cal S}_{q}(z)\big)^j\nn&&\oplus~\sum_{i',j'=1}^{q'}
{}~\sum_{k'=-\infty}^\infty\,\alpha_{i',j'+q'/2;k'}^{\prime}~z^{\prime\,k'}
\,\left({\cal C}_{q'}\right)^{i'}\,\big({\cal S}_{q'}(z'\,)\big)^{j'} \ ,
\label{Aexp2}\end{eqnarray}
where we assume that the matrix ranks $q$ and $q'$ are
even. Everything we say in the following will hold
symmetrically for both summands in (\ref{Aexp2}), and hence we will
only deal explicitly with the first one $\a(z)$. Its expansion
coefficients in (\ref{Aexp2}) may be computed from
the analogous coefficients in the expansion (\ref{Gammanhomo}) and we get
\be
\alpha_{i,j;k}=\sum_{l=-\infty}^\infty a_{i+(l-1/2)q\,,\,j+qk} \ .
\label{Aexp2Gamma}\end{equation} Because of the rapid decay of the
coefficients $a_{m,r}$ when $m$ is large, the information about
the high momentum modes of $U$ is lost. In the other summand one
would find that the loss is in the high momentum modes of $V$.
Since the sequences $\{a_{m,r}\}$ and $\{a'_{m,r}\}$ decrease rapidly, for $q$
and
$q'$ sufficiently large, the ``error'' is small. As  the two
summands neglect different high momentum modes, the two errors
``compensate'' each other. This provides a heuristic insight into
the role of the two summands in the matrix regularization.

Let us now look at the projection of $\del_1
a=2\pi\ii\,\sum_{m,r}m\,a_{m,r}~U^m\,V^r$ in the first summand. By using
(\ref{Aexp2Gamma}) the corresponding expansion coefficients may be
written as
\bea
\big(\gamma_n(\del_1 a)\big)_{i,j;k}&=&2\pi\ii\,\sum_{l=-\infty}^\infty
\left(i+(l-\mbox{$\frac12$})\,q\right)\,a_{i+(l-1/2)q\,,\,j+qk}\nn&=&
2\pi\ii\,\left(i-\mbox{$\frac q2$}\right)\,\alpha_{i,j;k}+O
\left(q\,a_{i-q/2\,,\,qk}\right) \ ,
\label{derivexpcoeffs}\end{eqnarray}
where the neglected terms in the second equality vanish for Schwartz
sequences as $n\to\infty$. The same reasoning can be repeated for
$\del_2$, and together these results suggest the definitions
\bea
\nabla_1\a(z)&=&2\pi\ii\,\sum_{i,j=1}^q~\sum_{k=-\infty}^\infty\,
  i\,\alpha_{i+q/2,j;k}~z^k\,\left({\cal C}_{q}\right)^i\,
\big({\cal S}_{q}(z)\big)^j ~, \nn
\nabla_2\a(z)&=&2\pi\ii\,\sum_{i,j=1}^q~\sum_{k=-\infty}^\infty\,
(j+q\,k)\,\alpha_{i+q/2,j;k}~z^k\,\left({\cal C}_{q}
\right)^i\,\big({\cal S}_{q}(z)\big)^j~.
\label{derin}\end{eqnarray}
These two operations converge to the canonical linear derivations
$\partial_\mu$, $\mu=1,2$ on the algebra $\atheta$ and satisfy an
approximate Leibniz rule~\cite{inprep}.

We will now express the ``derivatives'' (\ref{derin}) as
operations acting on $\a(z)$ expressed as a matrix-valued function
on a circle, as in (\ref{Az}). For this, we need to find the
appropriate change of matrix basis between the two expansions
(\ref{Aexp2}) and (\ref{Az}). The key to this is the identity,
readily derived from~(\ref{matrixmult}),
\be
\left({\cal C}_{q}\right)^i\,\big({\cal S}_{q}(z)\big)^j
=\sum_{s=1}^{q-j}\omega^{i(s-1)}~\P^{s,s+j}+z\, \sum_{s=q-j+1}^{q}
\omega^{i(s-1)}~\P^{s,s+j-q} \ . \label{keyidentity}\end{equation}
Using~(\ref{keyidentity}), the trace formula~(\ref{Pijtrace}), and
the fact that $\omega$ is a $q$-th root of unity, one can show
that the elements of the matrix basis of the
expansion~(\ref{Aexp2}) are orthogonal with respect to the usual
matrix trace. This, the identities~(\ref{keyidentity})
and~(\ref{Pijtrace}), and some algebra lead to the change of basis
$a_{ij;k}\mapsto \alpha_{i,j;k}$,
\be
\alpha_{i,j;k}=\frac1{q}\,\left[\,\sum_{s=1}^{q-j}a_{s,s+j;k}~
\omega^{-(i-q/2)(s-1)}+\sum_{s=q-j+1}^{q}a_{s,s+j-q;k+1}~
\omega^{-(i-q/2)(s-1)}\right] \ .\label{aAchange}\end{equation}
The inverse of the map (\ref{aAchange}) may be similarly computed
to get
\be
a_{ij;k}~=~\sum_{s=1}^{q}\,\omega^{s(i-1)}~\times~
\left\{\begin{matrix}\alpha_{s+q/2,j-i;k}&i<j\\
\,\alpha_{s+q/2,q+j-i;k+1}&i\geq j\end{matrix}\right. \ .
\label{Aachange}\end{equation}

Consider first  $\nabla_1$ acting in (\ref{derin}), and
substitute $\varepsilon_{i,j;k}:=(i-q/2)\,\alpha_{i,j;k}$ in place
of $\alpha_{i,j;k}$ in (\ref{Aachange}). For $i\leq j$ it then follows
from (\ref{aAchange}) that the corresponding canonical matrix
elements $e_{ij;k}^{(n)}$ of $\nabla_1\a(z)$ in the expansion (\ref{Az}) are
given by
\be
(\Sigma\a)_{ij;k}:=e_{ij;k}=\frac1{q}\,\sum_{s=1}^{q}s\,\omega^{si}
\,\left[\,\sum_{s'=1}^{q-j+i}a_{s',s'+j-i;k}~
\omega^{-ss'}+\sum_{s'=q+i-j+1}^{q}a_{s',s'+j-i-q;k+1}~
\omega^{-ss'}\right] \ , \label{bijkn}\end{equation}
with a similar expression in the case $i>j$. In matrix components its
action on matrix valued functions on a circle is given by
$\Sigma\a(\tau)_{ij}=\sum_{s,t}\Sigma(\tau)_{ij,st}\,\a(\tau)_{st}$
with
\be
\Sigma(\tau)_{ij,st}~=~-\frac{2\pi\ii}{q}\,\sum_{s'=1}^{q}
s'\,\omega^{s'(i-s)}~\times~\left\{\begin{matrix}
     \delta_{t,s+j-i}&i<j&1\leq s\leq q+i-j\\\,
     \delta_{t,s+j-i-q}~\e^{2\pi\ii\tau}
     &i<j&q+i-j+1\leq s\leq q\\\delta_{t,s+i-j}&j\leq
     i&1\leq s\leq q+j-i\\\delta_{t,s+i-j-q}~
     \e^{-2\pi\ii\tau}&j<i&q+j-i+1\leq s\leq q
   \end{matrix}\right. \ .
\label{SSigmaijkl}\end{equation} The skew-adjoint shift operator
$\Sigma$ defines the finite analog of the derivative $\partial_1$
acting on the matrix part of~(\ref{Az}). Proceeding to $\nabla_2$,
define the operator
\be
\Xi_{ij}:=2\pi\ii j~\delta_{ij} \label{Xidef} \ .
\end{equation}
Then we may write the canonical matrix expansion coefficients $f_{ij;k}$ of
$\zeta_{i,j;k}:=j\,\alpha_{i,j;k}$ in the expansion (\ref{Az}), using
(\ref{aAchange}) and (\ref{Aachange}), as
\be
f_{ij;k}=(j-i)\,a_{ij;k}=\frac1{2\pi\ii}\,
\big[\,\Xi\,,\,\a\,\big]_{ij;k} \ .
\label{cijkn}\end{equation}
The operator $\Xi$ defines the finite analog of
the derivative $\partial_2$ acting on the matrix part of the expansion
(\ref{Az}). The operators $\Xi$ and $\Sigma$ can be thought of,
respectively, as ``infinitesimal'' versions of the clock and shift
operations defined in (\ref{Sn0Dapprox},\ref{dmu0D}). In this
sense, the derivative terms in the ensuing matrix model actions will
resemble more closely those which are obtained from soliton expansion
on the noncommutative {\it plane}~\cite{R2soliton}, rather than those
obtained from lattice regularization.

Finally, the components of the derivation $\nabla_2$ in
(\ref{derin}) which are proportional to the circular Fourier
integers $k$ are evidently given by the derivative operators
$z\,\dd/\dd z$ of $\circles^1$ acting on $\a(z)$. Completely
analogous formulas hold also for the second summand of the finite
level matrix algebras. In this way we may represent the
approximate derivatives acting on matrix-valued functions on
$\circles^1$ as
\bea
\nabla_1(\a\oplus\a')(\tau,\tau'\,)&=&\Sigma\a(\tau)~\oplus~\big(q'~
\dot{\a}'(\tau'\,)+\Sigma'\a'(\tau'\,)\big) \ , \nn
\nabla_2(\a\oplus\a')(\tau,\tau'\,)&=&\left(q~\dot{\a}(\tau)+
\big[\Xi\,,\,\a(\tau)\big]\right)~\oplus~\big[\Xi'\,,\,
\a'(\tau'\,)\big] \ , \label{nablafinite}\end{eqnarray} where
$\dot{\a}(\tau):=\dd\a(\tau)/\dd\tau$ and
$\dot{\a}'(\tau'\,):=\dd\a'(\tau'\,)/\dd\tau'$.

We can now write down an action defined on elements of $\athetan$ and which
approximates well the action functional (\ref{NCstrfield}) as
\beq
S_n^{(1)}=\beta\,\int\limits_0^1\dd\tau~
\Tr\,{\cal L}\big[\mPhi(\tau)\,,\,\nabla_\mu\mPhi(\tau)\big]
+\beta'\,\int\limits_0^1
\dd\tau'~\Tr'\,{\cal L}\big[\mPhi'(\tau'\,)\,,\,\nabla_\mu
\mPhi'(\tau'\,)\big] \ ,
\label{NCstrfieldn}\end{equation}
where $\gamma_n(\Phi)=\mPhi(\tau)\oplus\mPhi'(\tau'\,)$. The
noncommutativity of the torus has now been transformed into matrix
noncommutativity. Note that the matrix model here is defined on a
\emph{sum} of two circles, and the procedure is exact in the limit, in
the sense that the matrix algebras $\athetan$ converge to
$\atheta$. The fact that (\ref{NCstrfieldn}) already involves
continuum fields is also the reason that the derivations
(\ref{nablafinite}) are infinitesimal versions of the usual lattice
ones in (\ref{Sn0Dapprox},\ref{dmu0D}), and in the present case the
removal of the matrix regularization does not require a complicated
double scaling limit involving a small lattice spacing parameter as
in~\cite{AMNS}. However, the large $n$ limit required is also not the
't~Hooft limit, as in the matrix models of~\cite{R2soliton}, because
the notion of planarity in the matrix model (\ref{NCstrfieldn}) is the
same as that in the original continuum noncommutative field
theory.

This matrix model regularization of noncommutative field theory has a
number of desirable features. For instance, perturbation theory at
finite $n$ is readily tractable by using the expansion (\ref{Aexp2})
in the orthogonal basis of clock and generalized shift matrices. From
(\ref{derin}) it follows that {\it both} kinetic and interaction terms
assume simple diagonal forms in this representation. Moreover, one can
show that the regulated field theory is free from UV/IR mixing~\cite{MvRS},
suggesting that such matrix models provide a good arena to explore
renormalization issues. It is tempting to speculate that the broken
symmetry phases observed in noncommutative scalar field
theories~\cite{stripes} could be due to the soliton expansions in the
finite level algebras $\athetan$ in terms of the projections and
partial isometries of the noncommutative torus, as these solitons
display themselves momentum non-conserving stripe patterns
(see figs.~\ref{zawinul3} and~\ref{zawinul4}). Dynamically, they may
be due to a Kosterlitz-Thouless type phase transition in the matrix
quantum mechanics which occurs in the large $n$ limit, whereby a
condensation of vortices in the vacuum is responsible for the breaking
of the translational symmetry. These matrix theories further provide simple
toy models of tachyon dynamics which reproduce quantitative properties
of D-branes in open string field theory. This fits in well with recent
matrix model computations of D-brane dynamics in two-dimensional
bosonic string theory~\cite{c1}. It is also possible to map
the matrix regularization defined by (\ref{Gammanhomo}) onto the
noncommutative plane in a way which regulates the standard GMS
solitons~\cite{GMS}. It thereby provides a means of regularizing field theories
on noncommutative $\IR^2$ by means of one-dimensional matrix
models. These applications will be described in detail in~\cite{inprep}.

\bigskip

\noindent{\bf Acknowledgments:} We thank G.~Elliott, D.~Evans,
J.-H.~Park and J.~Varilly for helpful discussions and
correspondence. The work of G.L. and F.L. was supported in part by the
{\sl Progetto di Ricerca di Interesse Nazionale SINTESI}. The work of
R.J.S. was supported in part by an Advanced Fellowship from the {\sl
  Particle Physics and Astronomy Research Council}~(U.K.). F.L. would
like to thank the Department of Mathematics at Heriot-Watt University
for hospitality during his sabbatical year.

\providecommand{\href}[2]{#2}

\end{document}